\documentclass{article}

\usepackage{prime}
\usepackage[utf8]{inputenc} 
\usepackage{amsfonts}       
\usepackage{nicefrac}       
\usepackage{graphicx}
\usepackage{float}
\usepackage{url}
\usepackage[export]{adjustbox}
\pdfoutput=1

\title{\textbf{A Deep Learning Approach for Semantic Segmentation of Unbalanced Data in Electron Tomography of Catalytic Materials}}

\author{
	Arda Genc\\
	Center for the Accelerated Maturation of Materials\\
	Department of Materials Science and Engineering\\
	The Ohio State University\\
	Columbus, OH \\
	\texttt{genc.2@osu.edu} \\
	\And
	Libor Kovarik$^{*}$\\
	Institute for Integrated Catalysis\\
	Pacific Northwest National Laboratory\\
	Richland, WA\\
	\texttt{libor.kovarik@pnnl.gov} \\
	\And
	Hamish L. Fraser\\
	Center for the Accelerated Maturation of Materials\\
	Department of Materials Science and Engineering\\
	The Ohio State University\\
	Columbus, OH \\
	\texttt{fraser.3@osu.edu} \\
}

\begin{document}
\maketitle

\begin{abstract}
	Heterogeneous catalysts possess complex surface and bulk structures, relatively poor intrinsic contrast, and often a sparse distribution of the catalytic nanoparticles (NPs), posing a significant challenge for image segmentation, including the current state-of-the-art deep learning methods. To tackle this problem, we apply a deep learning-based approach for the multi-class semantic segmentation of a $\gamma$-Alumina/Pt catalytic material in a class imbalance situation. Specifically, we used the weighted focal loss as a loss function and attached it to the U-Net's fully convolutional network architecture. We assessed the accuracy of our results using Dice similarity coefficient (DSC), recall, precision, and Hausdorff distance (HD) metrics on the overlap between the ground-truth and predicted segmentations. Our adopted U-Net model with the weighted focal loss function achieved an average DSC score of 0.96 $\pm$ 0.003 in the $\gamma$-Alumina support material and 0.84 $\pm$ 0.03 in the Pt NPs segmentation tasks. We report an average boundary-overlap error of less than 2 nm at the 90th percentile of HD for $\gamma$-Alumina and Pt NPs segmentations. The complex surface morphology of the $\gamma$-Alumina and its relation to the Pt NPs were visualized in 3D by the deep learning-assisted automatic segmentation of a large data set of high-angle annular dark-field (HAADF) scanning transmission electron microscopy (STEM) tomography reconstructions. 
	
\end{abstract}
\section{Introduction}
TEM tomography in materials science has become the de facto technique that enables valuable high spatial resolution information on the structure of the materials in 3D \cite{arslan_embedded_2005,midgley_3d_2003,midgley_nanoscale_2006,midgley_z-contrast_2001,weyland_electron_2001}. Despite the progress in developing novel methods in acquiring and aligning the tomography tilt series and a broad spectrum of reconstruction algorithms\cite{gursoy_rapid_2017,wang_07_2020,han_deep_2021,shepp_maximum_1982}, semantic segmentation of the large 3D data sets remains a significant bottleneck in 3D analysis. Manual segmentation is a very time-consuming task, relying heavily on the handcrafting skills and expertise of a human operator. A reliable, reproducible, and fully automated segmentation method is in high demand for scaling the 3D data analysis and collecting statically meaningful information where a single reconstructed volume segmentation involves hundreds of images or more. 

Recent advances in deep learning methods have revolutionized the field of computer vision\cite{lecun_deep_2015,krizhevsky_imagenet_2012,long_fully_2015,ciresan_deep_2012}, and the evolution of these methods has enabled the automatic semantic segmentation of large data sets; otherwise, manual analysis is unfeasible\cite{milletari_v-net_2016,decost_high_2019,dong_automatic_2017}. Deep learning pixel-wise classifiers have been successfully applied to many semantic segmentation tasks where complex structures are not easily mapped by simple intensity differences, and boundaries between the image features are not apparent due to the variations in contrast gradients\cite{greenspan_guest_2016}.
 
In deep learning, fully convolutional neural networks (FCNs) hierarchically recognize complex features directly from the training data without the additional feature engineering. More recently, FCNs, inspired by large and deep networks, are efficiently trained end-to-end by supervised learning and pixel-to-pixel probabilities computed successfully, thanks to the many advancements in parallel computing\cite{glorot_deep_2011}. It has been shown that segmentation results using FCNs can indeed reach a human-level performance and even on some occasions exceed that without the post-tuning of the results\cite{zeng_deepem3d_2017,greenwald_whole-cell_2021,roberts_deep_2019}. Today, it is possible to generalize these deep networks with a limited amount of ground-truth data by implementing data augmentation and regularization techniques while mitigating the problem of variance\cite{hernandez-garcia_data_2018}.

In this paper, we explore the capability of FCNs in the semantic segmentation of a $\gamma$-Alumina ($Al_2O_3$)/Pt catalytic material. Historically, $\gamma$-Alumina has been one of the most used catalytic support materials for noble metals and oxide catalysts employed for reduction, oxidation, and reforming reactions in automotive exhaust control and petroleum refining processes\cite{leach_applied_1983}. $\gamma$-Alumina possesses a complex crystalline structure; despite its well-known application space, the origin of the catalytic behavior is not well understood\cite{kovarik_tomography_2013,khivantsev_precise_2021}. There is considerable debate on the role of surfaces of $\gamma$-Alumina responsible for both catalytic behavior and anchoring of the noble metallic NPs. In addition to the structural complexity and small crystallite sizes, the support material $\gamma$-Alumina consists of a dense network of matrix pores, and the degree to which these pores are connected to outside surfaces is of great interest\cite{kovarik_tomography_2013,khivantsev_precise_2021,roiban_3d-tem_2012,epicier_2d_2019}.

In catalytic materials, the sparse distribution of the noble metallic NPs, over the background and oxide support material introduces an unbalanced representation of the data in TEM images. The extend of the class imbalance problem between the foreground and background of the images has been extensively studied in deep learning-based semantic segmentation approaches\cite{qin_weighted_2018,novikov_fully_2018,sugino_loss_2021}. It has been shown that the choice of loss function significantly impacts the performance of a semantic segmentation model\cite{yeung_unified_2021}. Many recent state-of-the-art applications of FCNs focus on the implementation of weighting strategies coupled with distribution-based or differentiable region-based loss functions for the optimization of the models\cite{yeung_unified_2021,jadon_survey_2020,sudre_generalised_2017}. 

Even though weighting strategies at the loss function level control the class imbalance, the problem of loss becoming overwhelmed by the number of easy examples during inference remains a challenge in complex multi-class situations\cite{lin_focal_2020}. Moreover, there are limited applications of these strategies in the semantic segmentation of materials science samples, particularly segmentation of the 3D electron tomography reconstructions\cite{decost_high_2019,roberts_deep_2019,horwath_understanding_2020,akers_rapid_2021}. 

We present a U-net-based FCN architecture and weighted focal loss as a loss function to overcome the class imbalance problem\cite{ronneberger_u-net_2015,lin_focal_2020}. The weighted focal loss is a distribution-based loss function, and weighting of the unbalanced data occurs at the loss function level in contrast to data preprocessing strategies. In addition to the weighting, focal loss applies a modulation term to the standard cross-entropy loss and dynamically scales the confidence of the correctly classified examples\cite{lin_focal_2020}. In our experiments, the U-Net architecture equipped with the weighted focal loss facilitated a comprehensive 3D representation of the catalytic material and provided a clear insight regarding the long-standing debate on the characteristics of $\gamma$-Alumina surfaces and their relation to the catalytic NPs. We discuss the accuracy of our segmentation results by assessing commonly used semantic segmentation metrics on the overall overlap and boundary match between the ground-truth and predicted segmentations. 

To further test our model’s robustness and validity, the best-performing model was applied to the automatic semantic segmentation of a large data set of reconstructed images. We believe strongly that deep learning-based semantic segmentation methods have immense potential in 3D data analysis and will usher in a new era in materials design and discovery. 

\section{Methods}
\subsection{3D tomography data acquisition and visualization}
Our main goal is to assess the effectiveness of a deep learning-based approach in semantic segmentation of the 3D HAADF STEM tomography reconstructions while achieving a full 3D view of the $\gamma$-Alumina/Pt catalytic material. For this purpose, we conducted HAADF STEM tomography experiments on a well-isolated $\gamma$-Alumina/Pt catalytic particle. TEM samples were prepared by dropping a solution containing well-dispersed NPs on a lacey carbon film. A detailed description of the synthesis of $\gamma$-Alumina/Pt material was reported in an earlier paper\cite{kovarik_tomography_2013}. A probe aberration-corrected 300 kV Thermo Fisher Scientific Titan S/TEM microscope was used for the acquisition of the HAADF STEM tilt series. HAADF STEM images were acquired at the detector inner collection angle of 40 mrad, beam current of 20 pA with a 0.1 nm probe size, an accelerating voltage of 200 kV. 

To extend the depth of focus of the electron beam during STEM tomography acquisitions, the convergence angle of the illumination system was adjusted to 10 mrad using the three-condenser lens optics of the microscope. Tomography tilt series consists of 69 HAADF STEM images acquired at the tilt range of $\pm$ 68$^{\circ}$ and tilt increment of 2$^{\circ}$.

Post-processing of the tilt series was conducted using open-source resources; a Python programming script based on the tomoviz software used for the image shift and tilt alignments\cite{noauthor_tomviz_nodate}, and TomoPy and ASTRA Toolbox Python libraries for the maximum likelihood expectation maximization (MLEM) reconstructions\cite{pelt_integration_2016,gursoy_tomopy_2014}. Paraview software was employed to generate 3D visualizations from the fully segmented 3D reconstructions\cite{ahrens_paraview_2005}.
\subsection{Segmentation architecture}
For FCNs experiments, we exploit an adopted version of the U-Net architecture, and a schematic of the architecture is shown in Figure 1. Our network consists of two learning paths, a down-sampling (contraction) path and an up-sampling (expansion) path. There are six convolutional steps in the down-sampling path and five in the up-sampling path. In the down-sampling path, each step has two convolutional layers with a filter size of 3x3. The size of the feature maps is halved by the pooling layers following each step. In the up-sampling path, each step starts with a convolutional transpose layer with a filter size of 2x2 and a stride of 2, followed by two convolutional layers with a filter size of 3x3. The size of the feature maps is doubled, and the number of feature maps is halved at the end of each convolutional step in the up-sampling path.

In our version of the U-Net, we used the ‘same’ padding in the convolution layers followed by an average pooling layer for down-sampling. Using the ‘same’ padding resulted in the output of the convolution layers being the same size as the input layers. The pooling operation plays a vital role in the flow of information through the convolutional layers and defines the model’s sensitivity to details. In our architecture, average pooling is preferred over max pooling for down-sampling to reduce the spatial information loss at the feature boundaries and prevent excessive pixel saturation. 

Concatenation paths (skip connections) give the network a well-known ‘U’ shape pattern and link the high spatial information from down-sampling convolutional layers to the up-sampling convolutional layers. The network hierarchically learns the contextual information and fine details in the predicted images. We did not see a significant performance improvement in our model with the addition of dropout layers; instead, we employed a comprehensive data augmentation strategy for regularization. We used a rectified linear unit (ReLU) activation function for hidden layers and softmax for the output convolutional layer with final feature maps of 3.

\subsection{Training and optimization}
For optimization, we used a mini-batch gradient descent with a batch size of 2 and Adam optimizer at a learning rate of 0.0005. We used default parameters from the Tensorflow deep learning framework for the first and second moments of gradient averaging and updating\cite{abadi_tensorflow_2016}. All the weights were initialized by "He normal" kernel initialization\cite{he_delving_2015}, and all the biases were initialized at 0. A total of 30 ground-truth images were selected from the 3D reconstructions, and corresponding ground-truth segmentations were manually annotated for the training and validation steps. A class label representing background/pores,  $\gamma$-Alumina, and Pt NPs, were assigned to each pixel in the ground-truth segmentations.  

Due to the large image size and limited GPU memory availability, 1024x512 pixels (0.12 nm/pixel) ground-truth images and segmentations were divided into 512x512 pixels patches. This data was then randomly split into 75\% training and 25\% validation data sets. The average pixel density of each class in the patches is 77.7\% for background/pores, 21.8\% for $\gamma$-Alumina, and 0.5\% for Pt NPs. 

Data augmentation is crucial to teach the network a robust invariance to input data and generalize the model. We used rotation, vertical and horizontal flip, zoom, and shear transformations during training to generate a diverse range of images representing variations in the location and shape of the features. The best-performing model was selected based on the evaluation performance and applied to the automatic segmentation of a stack of 702 3D reconstructions. We employed a smooth blending approach to form final predictions where 512x512 pixels size segmented patches were smoothly merged into 1024x512 pixels size final predictions using spline interpolations between the overlapping patches\cite{noauthor_make_2022}.

\begin{figure}[t]
	\centering
	\includegraphics[width=1\textwidth,trim=120 90 180 50, clip]{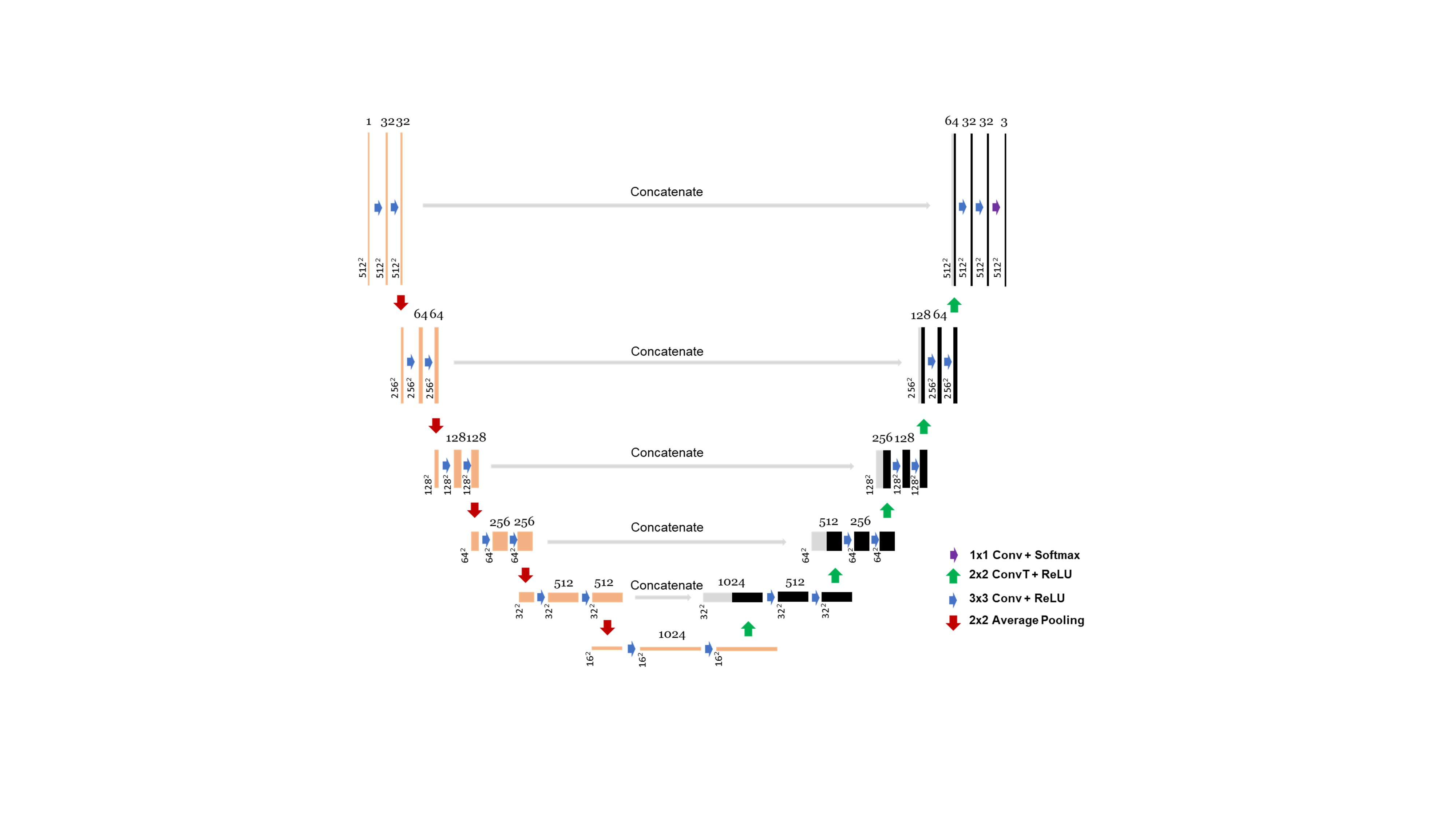}
	\caption{\small A schematic of the U-Net architecture. The number of feature maps is indicated on the top of each box, and the dimensions of the feature maps are on the bottom left corner. Orange boxes show the contraction path, and black boxes show the expansion path. The gray arrows and boxes indicate the concatenation path at each convolutional step.}
	\label{fig:fig1}
\end{figure}
As described previously, we investigate the complex bulk and surface structure of the $\gamma$-Alumina/Pt catalysts. Our segmentation model aims to distinguish the two-phase microstructure of the $\gamma$-Alumina and Pt NPs, as well as the pores. A correct classification of the surfaces, defined by the $\gamma$-Alumina - background and $\gamma$-Alumina – Pt - background boundaries, is mainly of high interest. In the tomography reconstructions, $\gamma$-Alumina and the background constitute the most significant fraction of the reconstructions, compared with the sparsely distributed nanoscale Pt particles. When there is an imbalance in the data representation, the learning algorithm can be biased towards the dominant class, represented densely by the higher number of pixels. In order to address this problem, we used weighted focal loss instead of the standard cross-entropy to minimize the loss\cite{lin_focal_2020,mavrin_focal_2022}.

Weighted focal loss is a differentiable modification
of the cross-entropy loss term and addresses the class imbalance problem in two ways. Firstly, it shifts the focus from easy-to-classify pixels towards hard-to-classify pixels by extending the range in which each pixel receives loss. This is achieved by scaling the cross-entropy loss by a focusing parameter and preventing the loss from being overwhelmed by the easy pixels. Secondly, the role of the weighted focal loss is to provide a balancing action on the class imbalance problem by adjusting the contribution of each class to the loss function by a weighting factor. Categorical focal loss ($L_{FL}$) is defined as the following in a multi-class problem:
\begin{equation}
	L_{F L}(y, p)=-\frac{1}{N} \sum_{i=1}^{N} \sum_{c=1}^{c} \alpha_{c} y_{i, c}\left(1-p_{i, c}\right)^{\gamma} \log p_{i, c}
\end{equation}
where $y_{(i,c)}$ and $p_{(i,c)}$ are the ground-truth and prediction probabilities of class c at pixel location i . Parameters C and N are the number of classes and pixels, respectively. $\alpha_c$ is the weighting factor for class c and $\gamma$ is the focusing parameter. Both focusing parameter and weighting factor are tunable hyperparameters. In our experiments, $\alpha_c$ values were approximated on the density of representation of each class at the range from 0 to 1, and $\gamma$ was set to 1.
\subsection{Evaluation metrics}
Evaluation metrics play an essential role in proving the network’s performance and thus establishing the model for automatic semantic segmentation. In this work, prediction results were assessed using four commonly used semantic segmentation metrics: Dice similarity coefficient (DSC), recall, precision, and Hausdorff distance (HD)\cite{taha_metrics_2015}. DSC, recall, and precision scores are similarly extracted from the confusion matrix and defined as:
\begin{equation}
	DSC=\frac{2TP}{2TP+FP+FN}
\end{equation}
\begin{equation}
	 Recall=\frac{TP}{TP+FN}
\end{equation}
\begin{equation}
	Precision =\frac{TP}{TP+FP}
\end{equation}
where true positives (TP), false positives (FP), and false negatives (FN) represent per pixel classifications of the confusion matrix.

We also evaluated our semantic segmentation results by measuring the dissimilarities specifically at the segmentation boundaries. Hausdorff distance is a boundary distance-based metric and measures the largest segmentation error in the overlap between the ground-truth and predicted segmentations\cite{huttenlocher_comparing_1993}. Given two sets of points A and B, Hausdorff distance is defined as:
\begin{equation}
	H D(A, B)=\max (h d(A, B), h d(B, A))
\end{equation}
where hd(A,B) and hd(B,A)  are directed Hausdorff distances:
\begin{equation}
	h d(A, B)=\max _{a \in A} \min _{b \in B}\|a-b\|
\end{equation}
\begin{equation}
	h d(B, A)=\max _{b \in B} \min _{a \in A}\|a-b\|
\end{equation}
Functions hd(A,B) and hd(B,A)  measure the distances between two points in A and B, which are farthest from any nearest neighbors, and HD (A,B) gives the largest of these distances. ‖a-b‖ is the Euclidean norm between the points in A and B. A well-documented behavior of the Hausdorff distance is its sensitivity to outliers and noise\cite{maiseli_hausdorff_2021}; thus, we report robust HD (RHD) values considering the percentile of the largest segmentation errors and as well as the maximum HD\cite{dong-gyu_sim_object_1999,noauthor_surface_2022}.  We aim to down-weight the impact of outliers and noise on the HD metric by measuring the RHD values.
\section{Results and discussion}
Segmentation of the HAADF STEM tomography reconstructions is a challenging task due to information loss from the insufficient number of projections (i.e., missing wedge artifacts) and variations in the contrast and size of the features in tilt images. A series of representative orthogonal slices (orthoslices) taken from the reconstructed 3D volume of an isolated $\gamma$-Alumina/Pt particle is shown in Figure 2. a-c. 
\begin{figure}[bh]
	\centering
	\includegraphics[width=0.5\textwidth,trim=300 50 270 50, clip]{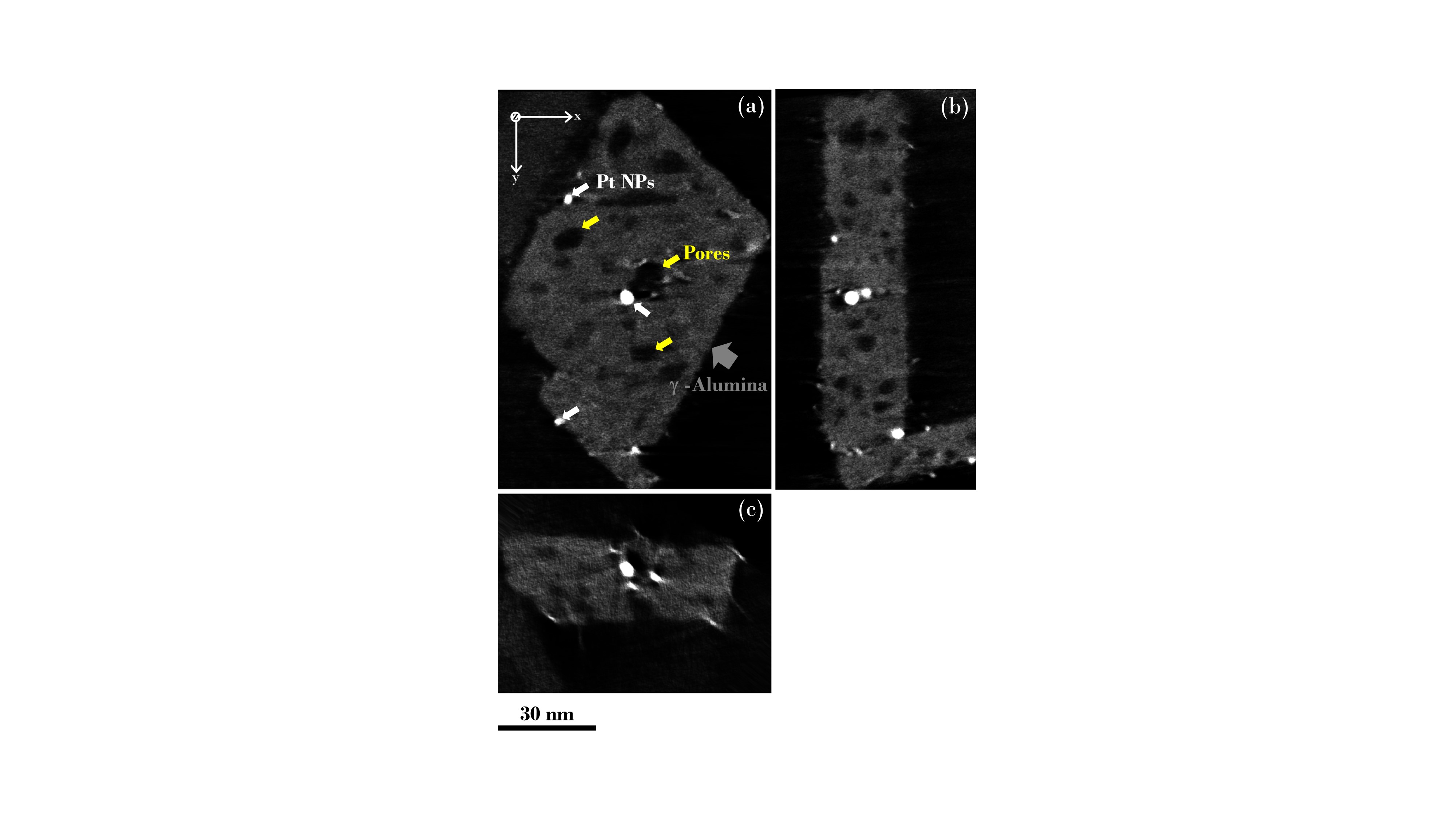}
	\caption{\small Orthogonal slices taken from the 3D HAADF STEM reconstructions of an isolated $\gamma$-Alumina/Pt particle.}
	\label{fig:fig2}
\end{figure}
\begin{figure}[t]
	\centering
	\includegraphics[width=1\textwidth,trim=100 10 100 10, clip]{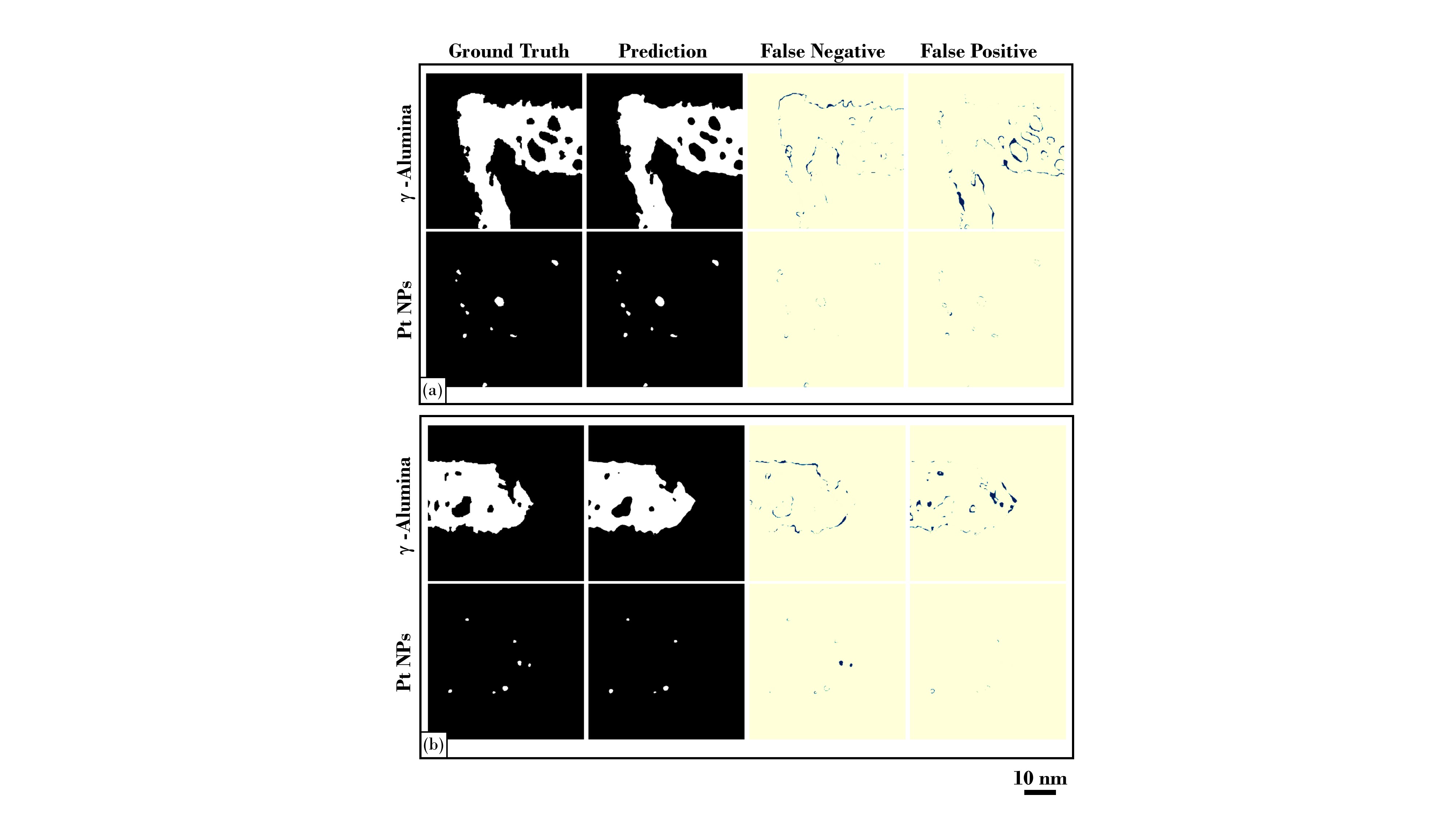}
	\caption{\small A set of ground-truth and predicted segmentations of $\gamma$-Alumina and Pt NPs from the validation data and corresponding false negative and false positive maps showing the discrepancies in the overlap of each class.}
	\label{fig:fig3}
\end{figure}
In the orthoslices, the 3D microstructure of the particle is sectioned perpendicular and parallel to the broad surfaces of the $\gamma$-Alumina. At first glance, we notice the significant contrast difference between the oxide $\gamma$-Alumina and metallic Pt NPs, where Pt NPs appear much brighter than both $\gamma$-Alumina substrate and background in the reconstructions. Pt NPs mainly exhibit round shapes, and their size distribution is in the 1- 4 nm range, while $\gamma$-Alumina particles have a thin plate-like structure and a roughly rhombus shape. Some of the very small Pt NPs show an elongated shape because of the well-documented missing wedge artifact in TEM tilt tomography\cite{midgley_3d_2003}, as seen in Figure 2.c. In $\gamma$-Alumina, the crystallographic shape of the particle is defined by the \{110\} and \{111\} type surfaces, and the main ‘broad’ surface is \{110\} orientation, and side surfaces are \{111\}. Furthermore, synthesized $\gamma$-Alumina particle shows small ledges and facets on the surfaces and a network of high-density pores inside the matrix.

To evaluate the performance of our model in the segmentation task, we compared the results obtained from the validation data. Figure 3. visualizes results of exemplar qualitative segmentations from $\gamma$-Alumina and Pt NPs, respectively. The images represent two 512x512pixels patches from the validation data. For each patch of the validation data, ground-truth and predicted segmentations of $\gamma$-Alumina and Pt NPs are compared separately in the binary images. The differences in the segmentations are shown more explicitly in the false negative and false positive maps highlighting the discrepancies in the overlap of each class. When considering the false positive maps of the $\gamma$-Alumina segmentations, we notice that $\gamma$-Alumina boundaries are relatively shifted towards the pores inside the $\gamma$-Alumina matrix. In contrast to $\gamma$-Alumina/pore boundaries, most misclassifications near the surfaces are discontinuous. False negative and false positive regions are extended only a few pixels wide towards either $\gamma$-Alumina or background. Moreover, the boundaries in the predicted segmentations, both on the surfaces and along the $\gamma$-Alumina/pore boundaries, are smoother than the ground-truth segmentations. 

We further discuss these observations in the context of the evaluation metrics, and the results are shown in Table 1. As expected, shifting of the $\gamma$-Alumina boundaries, mainly towards the pores, is reflected in the evaluation results such that the average precision score is lower than the average recall score for $\gamma$-Alumina. The average precision and recall scores are 0.95 $\pm$ 0.008 (mean $\pm$ $\%$95 confidence interval) and 0.97 $\pm$  0.004, respectively. 

In contrast to $\gamma$-Alumina, we observed fewer false positives in the segmentations of Pt NPs. Still, there are some missing Pt NPs in the predictions and corresponding misclassified regions seen in the comparison of the Pt NPs segmentations. Compared with $\gamma$-Alumina, the average precision score of Pt NPs is higher than the recall score. The average precision and recall scores are 0.92 $\pm$ 0.03 and 0.78 $\pm$ 0.04, respectively for the segmentation of Pt NPs. 

One explanation for the extension of the $\gamma$-Alumina matrix towards the pores inside may be the contrast modulations along the diffuse boundaries between the $\gamma$-Alumina and pores. Uncertainty in the contrast of these boundaries can potentially fuel ambiguity in the manual annotations of the ground-truth segmentations. Still, a precise annotation of these low contrast boundaries can be a challenging task even for a human expert. Another approach would be generating more annotated data to elevate the model’s performance. However, this would be computationally costly and would require additional manual annotations. It is also worth mentioning that it is inevitable to have false negatives and false positives during inference. Practically, we would aim a trade-off between precision and recall. 

Nevertheless, a comparison of the ground-truth segmentations with the predicted segmentations shows a strong correlation, especially in the complex surface structure of the $\gamma$-Alumina and the appearance of the Pt NPs. The overall similarities in the size and shape of the Pt NPs between the ground-truth and predicted segmentations also suggest that the model has convincingly managed the class imbalance problem without a significant underestimation. The overall segmentation performance measured by the DSC score for each class is 0.99 $\pm$ 0.002 for background/pores, 0.96 $\pm$ 0.003 for $\gamma$-Alumina, and 0.84 $\pm$ 0.03 for Pt NPs. Here, we report the background DSC score since the pores inside $\gamma$-Alumina are associated with the background class.\\
\setlength{\tabcolsep}{10pt} 
\renewcommand{\arraystretch}{1.5} 
\begin{table}
	\centering
	\caption{Evaluation results from the validation data. All the values are in the form of mean $\pm$ \%95 confidence interval.}
	\begin{tabular}{cccc}
		Evaluation Metrics & $\gamma$-Alumina & Pt NPs & Background\textbackslash Pores \\
		\hline
		Precision &  0.95 $\pm$ 0.008 & 0.92 $\pm$ 0.03 & 0.99 $\pm$ 0.001 \\
		Recall &  0.97 $\pm$ 0.004 & 0.78 $\pm$ 0.04 & 0.99 $\pm$ 0.003 \\
		DSC &  0.96 $\pm$ 0.003 & 0.84 $\pm$ 0.03 & 0.99 $\pm$ 0.002 \\
		\\
	\end{tabular}
	\label{tab:table1}
\end{table}
\begin{figure}
	\centering
	\includegraphics[width=0.55\textwidth]{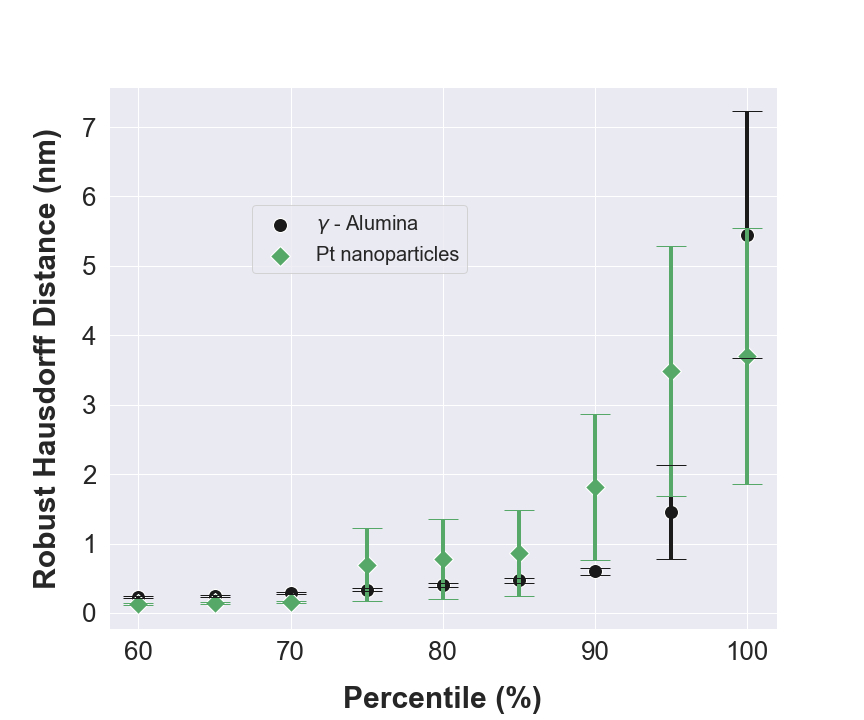}
	\caption{\small A plot of robust Hausdorff distance (RHD) vs. percentile of the largest segmentation error. Data points are in the form of mean $\pm$ standard error.}
	\label{fig:fig4}
\end{figure}\\
To investigate our segmentation results further, we conducted measurements on the degree of boundary match using the HD metric. This is of particular interest for detecting the model’s performance, correctly learning the boundaries in each class. The HD metric is a powerful tool in measuring the largest segmentation error in the overlap between the two segmentations, while the DSC score is an overlap metric affected by the segmentation performance over the entire image. Figure 4. shows the trend in the average HD of  $\gamma$-Alumina and Pt NPs segmentations at the various percentiles of robust HD (RHD) values. Measurement of the RHD provides a unique opportunity to understand the contributions of the outliers and noise to the model performance while guiding the degree of boundary match.
As seen in Figure 4., RHD values fall sharply from a maximum HD of 5.45 $\pm$ 1.78 nm (mean $\pm$ standard error) for $\gamma$-Alumina and 3.70 $\pm$ 1.84 nm for Pt NPs. At RHD95, HD between the ground-truth and predicted segmentations decreases to 1.45 $\pm$ 0.68 nm for $\gamma$-Alumina and 3.48 $\pm$ 1.80 nm for Pt NPs, and at RHD90, the largest segmentation errors are less than 2 nm (0.59 $\pm$ 0.05 nm for $\gamma$-Alumina and 1.81 $\pm$ 1.05 nm for Pt NPs). The variations in the RHD values show that the largest segmentation error for the Pt NPs trend higher than for the $\gamma$-Alumina, which is consistent with the overall lower evaluation scores observed for the Pt NPs, particularly the lower average recall score. Yet, this analysis suggests that our model learned reasonably well to classify the pixels at the boundaries, especially those of the $\gamma$-Alumina in addition to the matrix regions. 
\begin{figure}[t]
	\centering
	\includegraphics[width=1\textwidth,trim=150 100 150 60, clip]{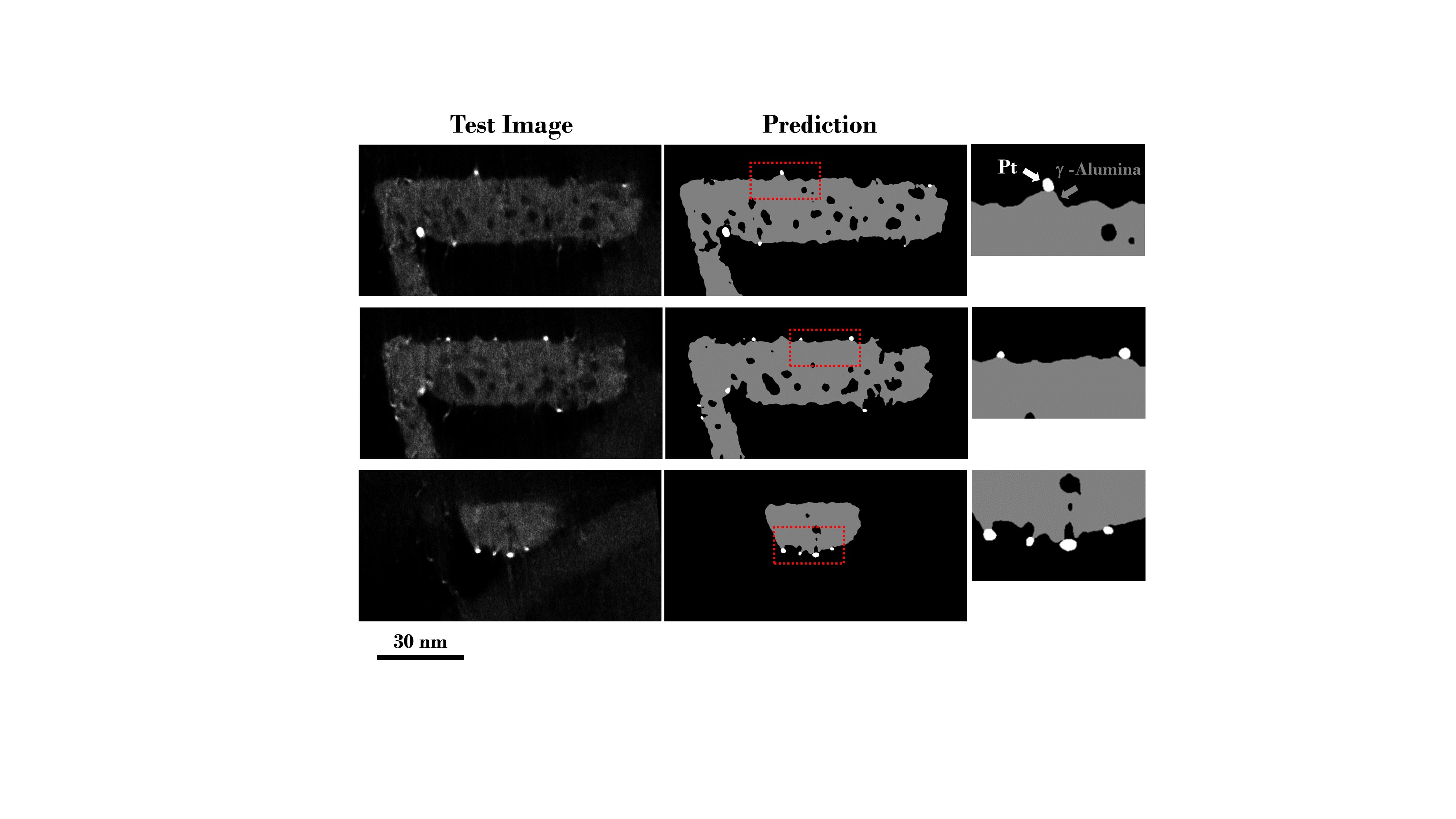}
	\caption{\small Comparison of the test images (3D reconstructions) and predicted segmentations of $\gamma$-Alumina catalytic support material and Pt NPs.}
	\label{fig:fig5}
\end{figure}

Based on the evaluation results, we applied the best-performing model for the automatic semantic segmentation of a large volume of 3D reconstructions. Figure 5. compares the example reconstructions extracted at various locations along the 3D volume of the $\gamma$-Alumina/Pt particle and corresponding predicted segmentations by the model. These reconstructions are test images, and the model has not seen them during the training and validation steps. In the predicted segmentations, the pixels classified as $\gamma$-Alumina are denoted in gray, Pt in white, and background/pores in black. 

Overall, there is a good correspondence between the test images and predicted segmentations. The location, shape, and size of the Pt NPs correlate with the test images and the texture of the pores inside the $\gamma$-Alumina matrix. One striking observation is that the catalytic Pt NPs are associated with the crystallographic modulations on the broad \{110\} surfaces of the $\gamma$-Alumina particle. Most of the Pt NPs are found at the apex of the two \{111\} type facets, as visualized explicitly in the zoomed-in images. 

Our model aims to segment accurately the complex surface and bulk microstructure of the $\gamma$-Alumina particle and Pt NPs with a limited amount of annotated ground-truth data. The results presented establish that the U-Net model with a weighted focal loss provides a stable model for the multi-class semantic segmentation of a large data set of 3D reconstructions in a severe class imbalance situation. 

Segmentation results establish a basis for the quantification of critical microstructural parameters, including quantification of the external surfaces in terms of their general area and proportion of the individual facets, quantification of the volume fraction of pores, and their surface area, Pt particle attachment, etc. While this topic will be the focus of our future work, an essential qualitative assessment of the $\gamma$-Alumina surfaces and geometry of the Pt NPs, can be obtained by transforming the stack of predicted segmentations into 3D visualizations. 

Figure 6. shows 3D volume visualization of $\gamma$-Alumina and Pt NPs from different viewpoints, and surface contour maps in light gray represent  $\gamma$-Alumina and red Pt NPs. 3D visualizations show that \{110\} surfaces of $\gamma$-Alumina are not atomically flat; instead, they form a series of periodically repeating structural facets. These facets are mostly terminated towards the center of $\gamma$-Alumina, and Pt NPs are anchored along with the \{111\} type facets rather than randomly distributed on the surfaces. Surprisingly, matrix pores are aligned along the direction of the surface facets, as seen in Figure 6.d.

\begin{figure}[t]
	\centering
	\includegraphics[width=1\textwidth,trim=100 10 100 10, clip]{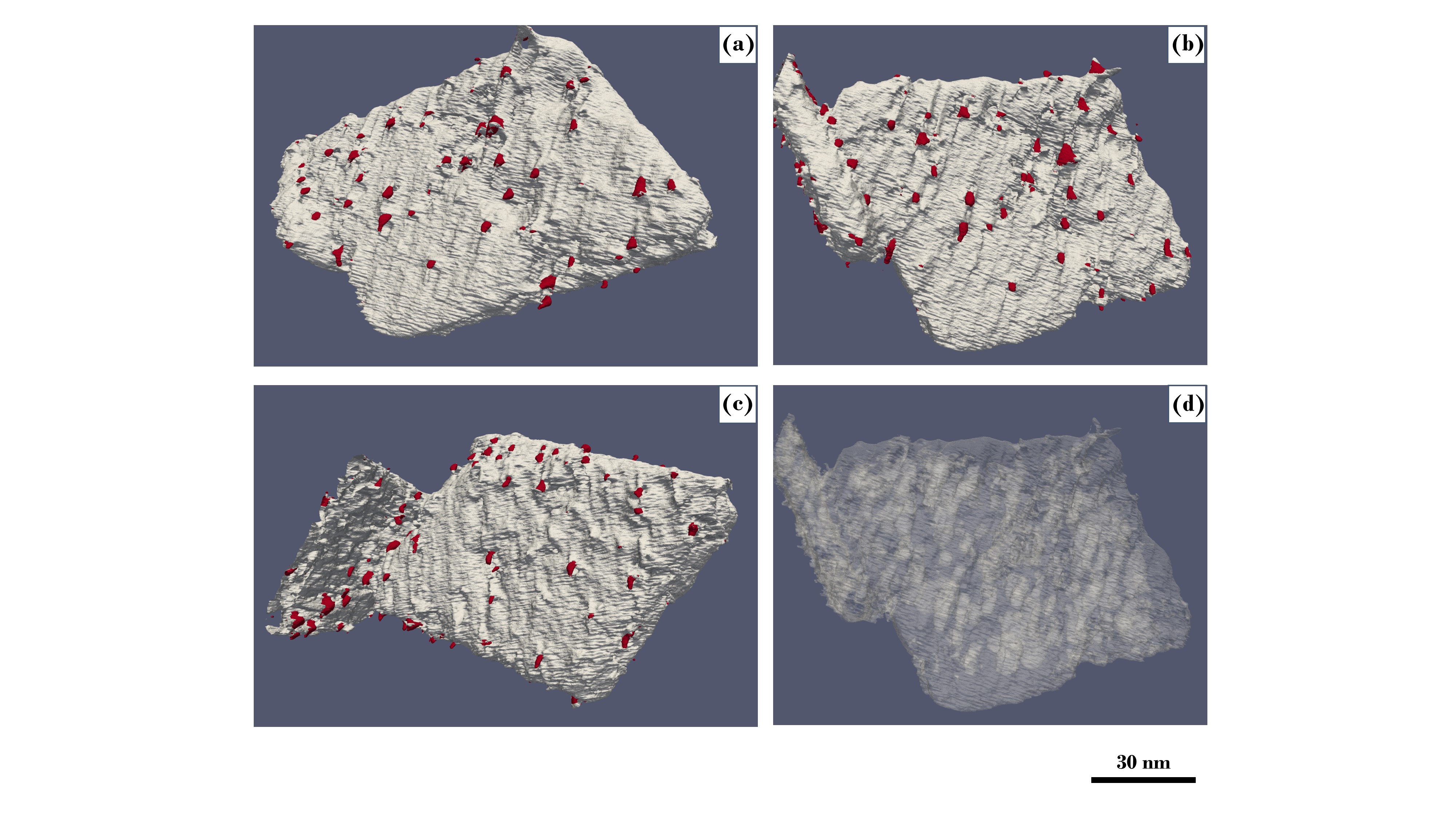}
	\caption{\small 3D visualization of $\gamma$-Alumina/Pt catalytic particle from different viewpoints.}
	\label{fig:fig6}
\end{figure}

\section{Conclusion}
We have demonstrated the effectiveness of a deep learning model in multi-class semantic segmentation of large and unbalanced data. This work was in many respects exploratory as a proof of concept, with a focus mainly on model performance. As such, the current model is naturally limited in extending its applicability. Alumina-based catalysts with various morphologies and particle contrasts will require additional adaptations to the current model. We intend to continue to utilize this U-net model in a transfer learning environment, incorporating the broad morphological and size variation of Alumina-based catalysts and eventually with general catalyst systems to achieve wider applicability. With recent advances in automatic data collection, deep learning-assisted semantic segmentation is genuinely expected to broaden the field of STEM tomography for routine quantitative measurement of catalysts on a statistically relevant scale, which is not possible currently.

\section*{Acknowledgement}
The experimental work was conducted in the Environmental Molecular Sciences Laboratory (EMSL), a national scientific user facility sponsored by the Department of Energy’s Office of Biological and Environmental Research at Pacific Northwest National Laboratory (PNNL). PNNL is a multi-program national laboratory operated for the DOE by Battelle Memorial Institute under Contract DE-AC05- 76RL01830. LK was supported by the U.S. Department of Energy (DOE), Office of Basic Energy Sciences, Division of Chemical Sciences, Geosciences, and Biosciences.

\bibliographystyle{ieeetr}
\bibliography{alumina_paper_original}

\begin{thebibliography}{10}

\bibitem{arslan_embedded_2005}
I.~Arslan, T.~J.~V. Yates, N.~D. Browning, and P.~A. Midgley, ``Embedded
  {Nanostructures} {Revealed} in {Three} {Dimensions},'' {\em Science},
  vol.~309, pp.~2195--2198, Sept. 2005.

\bibitem{midgley_3d_2003}
P.~Midgley and M.~Weyland, ``{3D} electron microscopy in the physical sciences:
  the development of {Z}-contrast and {EFTEM} tomography,'' {\em
  Ultramicroscopy}, vol.~96, pp.~413--431, Sept. 2003.

\bibitem{midgley_nanoscale_2006}
P.~A. Midgley, M.~Weyland, T.~J.~V. Yates, I.~Arslan, R.~E. Dunin-Borkowski,
  and J.~M. Thomas, ``Nanoscale scanning transmission electron tomography,''
  {\em Journal of Microscopy}, vol.~223, pp.~185--190, Sept. 2006.

\bibitem{midgley_z-contrast_2001}
P.~A. Midgley, M.~Weyland, J.~M. Thomas, and B.~F.~G. Johnson, ``Z-{Contrast}
  tomography: a technique in three-dimensional nanostructural analysis based on
  {Rutherford} scattering,'' {\em Chemical Communications}, no.~10,
  pp.~907--908, 2001.

\bibitem{weyland_electron_2001}
M.~Weyland, P.~A. Midgley, and J.~M. Thomas, ``Electron {Tomography} of
  {Nanoparticle} {Catalysts} on {Porous} {Supports}: {A} {New} {Technique}
  {Based} on {Rutherford} {Scattering},'' {\em The Journal of Physical
  Chemistry B}, vol.~105, pp.~7882--7886, Aug. 2001.

\bibitem{gursoy_rapid_2017}
D.~Gürsoy, Y.~P. Hong, K.~He, K.~Hujsak, S.~Yoo, S.~Chen, Y.~Li, M.~Ge, L.~M.
  Miller, Y.~S. Chu, V.~De~Andrade, K.~He, O.~Cossairt, A.~K. Katsaggelos, and
  C.~Jacobsen, ``Rapid alignment of nanotomography data using joint iterative
  reconstruction and reprojection,'' {\em Scientific Reports}, vol.~7,
  p.~11818, Dec. 2017.

\bibitem{wang_07_2020}
C.~Wang, G.~Ding, Y.~Liu, and H.~L. Xin, ``0.7 Å {Resolution} {Electron}
  {Tomography} {Enabled} by {Deep}‐{Learning}‐{Aided} {Information}
  {Recovery},'' {\em Advanced Intelligent Systems}, vol.~2, p.~2000152, Dec.
  2020.

\bibitem{han_deep_2021}
Y.~Han, J.~Jang, E.~Cha, J.~Lee, H.~Chung, M.~Jeong, T.-G. Kim, B.~G. Chae,
  H.~G. Kim, S.~Jun, S.~Hwang, E.~Lee, and J.~C. Ye, ``Deep learning
  {STEM}-{EDX} tomography of nanocrystals,'' {\em Nature Machine Intelligence},
  vol.~3, pp.~267--274, Mar. 2021.

\bibitem{shepp_maximum_1982}
L.~A. Shepp and Y.~Vardi, ``Maximum {Likelihood} {Reconstruction} for
  {Emission} {Tomography},'' {\em IEEE Transactions on Medical Imaging},
  vol.~1, pp.~113--122, Oct. 1982.

\bibitem{lecun_deep_2015}
Y.~LeCun, Y.~Bengio, and G.~Hinton, ``Deep learning,'' {\em Nature}, vol.~521,
  pp.~436--444, May 2015.

\bibitem{krizhevsky_imagenet_2012}
A.~Krizhevsky, I.~Sutskever, and G.~E. Hinton, ``{ImageNet} {Classification}
  with {Deep} {Convolutional} {Neural} {Networks},'' in {\em Advances in
  {Neural} {Information} {Processing} {Systems}} (F.~Pereira, C.~J.~C. Burges,
  L.~Bottou, and K.~Q. Weinberger, eds.), vol.~25, Curran Associates, Inc.,
  2012.

\bibitem{long_fully_2015}
J.~Long, E.~Shelhamer, and T.~Darrell, ``Fully convolutional networks for
  semantic segmentation,'' in {\em 2015 {IEEE} {Conference} on {Computer}
  {Vision} and {Pattern} {Recognition} ({CVPR})}, (Boston, MA, USA),
  pp.~3431--3440, IEEE, June 2015.

\bibitem{ciresan_deep_2012}
D.~Ciresan, A.~Giusti, L.~Gambardella, and J.~Schmidhuber, ``Deep {Neural}
  {Networks} {Segment} {Neuronal} {Membranes} in {Electron} {Microscopy}
  {Images},'' in {\em Advances in {Neural} {Information} {Processing}
  {Systems}} (F.~Pereira, C.~J.~C. Burges, L.~Bottou, and K.~Q. Weinberger,
  eds.), vol.~25, Curran Associates, Inc., 2012.

\bibitem{milletari_v-net_2016}
F.~Milletari, N.~Navab, and S.-A. Ahmadi, ``V-{Net}: {Fully} {Convolutional}
  {Neural} {Networks} for {Volumetric} {Medical} {Image} {Segmentation},'' in
  {\em 2016 {Fourth} {International} {Conference} on {3D} {Vision} ({3DV})},
  (Stanford, CA, USA), pp.~565--571, IEEE, Oct. 2016.

\bibitem{decost_high_2019}
B.~L. DeCost, B.~Lei, T.~Francis, and E.~A. Holm, ``High {Throughput}
  {Quantitative} {Metallography} for {Complex} {Microstructures} {Using} {Deep}
  {Learning}: {A} {Case} {Study} in {Ultrahigh} {Carbon} {Steel},'' {\em
  Microscopy and Microanalysis}, vol.~25, pp.~21--29, Feb. 2019.

\bibitem{dong_automatic_2017}
H.~Dong, G.~Yang, F.~Liu, Y.~Mo, and Y.~Guo, ``Automatic {Brain} {Tumor}
  {Detection} and {Segmentation} {Using} {U}-{Net} {Based} {Fully}
  {Convolutional} {Networks},'' {\em ArXiv}, vol.~abs/1705.03820, 2017.

\bibitem{greenspan_guest_2016}
H.~Greenspan, B.~van Ginneken, and R.~M. Summers, ``Guest {Editorial} {Deep}
  {Learning} in {Medical} {Imaging}: {Overview} and {Future} {Promise} of an
  {Exciting} {New} {Technique},'' {\em IEEE Transactions on Medical Imaging},
  vol.~35, no.~5, pp.~1153--1159, 2016.

\bibitem{glorot_deep_2011}
X.~Glorot, A.~Bordes, and Y.~Bengio, ``Deep {Sparse} {Rectifier} {Neural}
  {Networks},'' in {\em Proceedings of the {Fourteenth} {International}
  {Conference} on {Artificial} {Intelligence} and {Statistics}} (G.~Gordon,
  D.~Dunson, and M.~Dudík, eds.), vol.~15 of {\em Proceedings of {Machine}
  {Learning} {Research}}, (Fort Lauderdale, FL, USA), pp.~315--323, PMLR, Apr.
  2011.

\bibitem{zeng_deepem3d_2017}
T.~Zeng, B.~Wu, and S.~Ji, ``{DeepEM3D}: approaching human-level performance on
  {3D} anisotropic {EM} image segmentation,'' {\em Bioinformatics}, vol.~33,
  pp.~2555--2562, Mar. 2017.
\newblock \_eprint:
  https://academic.oup.com/bioinformatics/article-pdf/33/16/2555/25163663/btx188.pdf.

\bibitem{greenwald_whole-cell_2021}
N.~F. Greenwald, G.~Miller, E.~Moen, A.~Kong, A.~Kagel, T.~Dougherty, C.~C.
  Fullaway, B.~J. McIntosh, K.~X. Leow, M.~S. Schwartz, C.~Pavelchek, S.~Cui,
  I.~Camplisson, O.~Bar-Tal, J.~Singh, M.~Fong, G.~Chaudhry, Z.~Abraham,
  J.~Moseley, S.~Warshawsky, E.~Soon, S.~Greenbaum, T.~Risom, T.~Hollmann,
  S.~C. Bendall, L.~Keren, W.~Graf, M.~Angelo, and D.~Van~Valen, ``Whole-cell
  segmentation of tissue images with human-level performance using large-scale
  data annotation and deep learning,'' {\em Nature Biotechnology}, Nov. 2021.

\bibitem{roberts_deep_2019}
G.~Roberts, S.~Y. Haile, R.~Sainju, D.~J. Edwards, B.~Hutchinson, and Y.~Zhu,
  ``Deep {Learning} for {Semantic} {Segmentation} of {Defects} in {Advanced}
  {STEM} {Images} of {Steels},'' {\em Scientific Reports}, vol.~9, p.~12744,
  Sept. 2019.

\bibitem{hernandez-garcia_data_2018}
A.~Hernández-García and P.~König, ``Data augmentation instead of explicit
  regularization,'' {\em ArXiv}, vol.~abs/1806.03852, 2018.

\bibitem{leach_applied_1983}
B.~E. Leach, {\em Applied industrial catalysis {Volume} 1 {Volume} 1}.
\newblock New York: Academic Press, 1983.
\newblock OCLC: 841170071.

\bibitem{kovarik_tomography_2013}
L.~Kovarik, A.~Genc, C.~Wang, A.~Qiu, C.~H.~F. Peden, J.~Szanyi, and J.~H.
  Kwak, ``Tomography and {High}-{Resolution} {Electron} {Microscopy} {Study} of
  {Surfaces} and {Porosity} in a {Plate}-like gamma-{Al2O3},'' {\em The Journal
  of Physical Chemistry C}, vol.~117, pp.~179--186, Jan. 2013.
\newblock Publisher: American Chemical Society.

\bibitem{khivantsev_precise_2021}
K.~Khivantsev, N.~R. Jaegers, J.~Kwak, J.~Szanyi, and L.~Kovarik, ``Precise
  {Identification} and {Characterization} of {Catalytically} {Active} {Sites}
  on the {Surface} of gamma-{Alumina},'' {\em Angewandte Chemie}, vol.~133,
  pp.~17663--17671, Aug. 2021.

\bibitem{roiban_3d-tem_2012}
L.~Roiban, L.~Sorbier, C.~Pichon, C.~Pham-Huu, M.~Drillon, and O.~Ersen,
  ``{3D}-{TEM} investigation of the nanostructure of a delta-{Al2O3} catalyst
  support decorated with {Pd} nanoparticles,'' {\em Nanoscale}, vol.~4, no.~3,
  pp.~946--954, 2012.
\newblock Publisher: The Royal Society of Chemistry.

\bibitem{epicier_2d_2019}
T.~Epicier, S.~Koneti, P.~Avenier, A.~Cabiac, A.-S. Gay, and L.~Roiban, ``{2D}
  \& {3D} in situ study of the calcination of {Pd} nanocatalysts supported on
  delta-{Alumina} in an {Environmental} {Transmission} {Electron}
  {Microscope},'' {\em Catalysis Today}, vol.~334, pp.~68--78, 2019.

\bibitem{qin_weighted_2018}
R.~Qin, K.~Qiao, L.~Wang, L.~Zeng, J.~Chen, and B.~Yan, ``Weighted {Focal}
  {Loss}: {An} {Effective} {Loss} {Function} to {Overcome} {Unbalance}
  {Problem} of {Chest} {X}-ray14,'' {\em IOP Conference Series: Materials
  Science and Engineering}, vol.~428, p.~012022, Oct. 2018.
\newblock Publisher: IOP Publishing.

\bibitem{novikov_fully_2018}
A.~A. Novikov, D.~Lenis, D.~Major, J.~Hlaeovka, M.~Wimmer, and K.~Bühler,
  ``Fully {Convolutional} {Architectures} for {Multiclass} {Segmentation} in
  {Chest} {Radiographs},'' {\em IEEE Transactions on Medical Imaging}, vol.~37,
  pp.~1865--1876, 2018.

\bibitem{sugino_loss_2021}
T.~Sugino, T.~Kawase, S.~Onogi, T.~Kin, N.~Saito, and Y.~Nakajima, ``Loss
  {Weightings} for {Improving} {Imbalanced} {Brain} {Structure} {Segmentation}
  {Using} {Fully} {Convolutional} {Networks},'' {\em Healthcare}, vol.~9,
  no.~8, 2021.

\bibitem{yeung_unified_2021}
M.~Yeung, E.~Sala, C.-B. Schönlieb, and L.~Rundo, ``Unified {Focal} loss:
  {Generalising} {Dice} and cross entropy-based losses to handle class
  imbalanced medical image segmentation.,'' {\em Computerized medical imaging
  and graphics : the official journal of the Computerized Medical Imaging
  Society}, vol.~95, p.~102026, 2021.

\bibitem{jadon_survey_2020}
S.~Jadon, ``A survey of loss functions for semantic segmentation,'' {\em 2020
  IEEE Conference on Computational Intelligence in Bioinformatics and
  Computational Biology (CIBCB)}, pp.~1--7, 2020.

\bibitem{sudre_generalised_2017}
C.~H. Sudre, W.~Li, T.~K.~M. Vercauteren, S.~Ourselin, and M.~J. Cardoso,
  ``Generalised {Dice} overlap as a deep learning loss function for highly
  unbalanced segmentations,'' {\em Deep learning in medical image analysis and
  multimodal learning for clinical decision support : Third International
  Workshop, DLMIA 2017, and 7th International Workshop, ML-CDS 2017, held in
  conjunction with MICCAI 2017 Quebec City, QC,...}, vol.~2017, pp.~240--248,
  2017.

\bibitem{lin_focal_2020}
T.-Y. Lin, P.~Goyal, R.~Girshick, K.~He, and P.~Dollár, ``Focal {Loss} for
  {Dense} {Object} {Detection},'' {\em IEEE Transactions on Pattern Analysis
  and Machine Intelligence}, vol.~42, no.~2, pp.~318--327, 2020.

\bibitem{horwath_understanding_2020}
J.~P. Horwath, D.~N. Zakharov, R.~Mégret, and E.~A. Stach, ``Understanding
  important features of deep learning models for segmentation of
  high-resolution transmission electron microscopy images,'' {\em npj
  Computational Materials}, vol.~6, p.~108, Dec. 2020.

\bibitem{akers_rapid_2021}
S.~Akers, E.~Kautz, A.~Trevino-Gavito, M.~Olszta, B.~E. Matthews, L.~Wang,
  Y.~Du, and S.~R. Spurgeon, ``Rapid and flexible segmentation of electron
  microscopy data using few-shot machine learning,'' {\em npj Computational
  Materials}, vol.~7, p.~187, Dec. 2021.

\bibitem{ronneberger_u-net_2015}
O.~Ronneberger, P.~Fischer, and T.~Brox, ``U-{Net}: {Convolutional} {Networks}
  for {Biomedical} {Image} {Segmentation},'' in {\em {MICCAI}}, 2015.

\bibitem{noauthor_tomviz_nodate}
``tomviz for tomographic visualization of nanoscale materials.''
  \url{https://tomviz.org/}.

\bibitem{pelt_integration_2016}
D.~M. Pelt, D.~Gürsoy, W.~J. Palenstijn, J.~Sijbers, F.~De~Carlo, and K.~J.
  Batenburg, ``Integration of {TomoPy} and the {ASTRA} toolbox for advanced
  processing and reconstruction of tomographic synchrotron data,'' {\em Journal
  of Synchrotron Radiation}, vol.~23, pp.~842--849, May 2016.

\bibitem{gursoy_tomopy_2014}
D.~Gürsoy, F.~De~Carlo, X.~Xiao, and C.~Jacobsen, ``{TomoPy}: a framework for
  the analysis of synchrotron tomographic data,'' {\em Journal of synchrotron
  radiation}, vol.~21, pp.~1188--1193, Sept. 2014.
\newblock Edition: 2014/08/01 Publisher: International Union of
  Crystallography.

\bibitem{ahrens_paraview_2005}
J.~P. Ahrens, B.~Geveci, and C.~C. Law, ``{ParaView}: {An} {End}-{User} {Tool}
  for {Large}-{Data} {Visualization},'' in {\em The {Visualization}
  {Handbook}}, 2005.

\bibitem{abadi_tensorflow_2016}
M.~Abadi, A.~Agarwal, P.~Barham, E.~Brevdo, Z.~Chen, C.~Citro, G.~S. Corrado,
  A.~Davis, J.~Dean, M.~Devin, S.~Ghemawat, I.~J. Goodfellow, A.~Harp,
  G.~Irving, M.~Isard, Y.~Jia, R.~Józefowicz, L.~Kaiser, M.~Kudlur,
  J.~Levenberg, D.~Mané, R.~Monga, S.~Moore, D.~G. Murray, C.~Olah,
  M.~Schuster, J.~Shlens, B.~Steiner, I.~Sutskever, K.~Talwar, P.~A. Tucker,
  V.~Vanhoucke, V.~Vasudevan, F.~B. Viégas, O.~Vinyals, P.~Warden,
  M.~Wattenberg, M.~Wicke, Y.~Yu, and X.~Zheng, ``{TensorFlow}: {Large}-{Scale}
  {Machine} {Learning} on {Heterogeneous} {Distributed} {Systems},'' {\em
  ArXiv}, vol.~abs/1603.04467, 2016.

\bibitem{he_delving_2015}
K.~He, X.~Zhang, S.~Ren, and J.~Sun, ``Delving {Deep} into {Rectifiers}:
  {Surpassing} {Human}-{Level} {Performance} on {ImageNet} {Classification},''
  in {\em 2015 {IEEE} {International} {Conference} on {Computer} {Vision}
  ({ICCV})}, pp.~1026--1034, 2015.

\bibitem{noauthor_make_2022}
``Make smooth predictions by blending image patches, such as for image
  segmentation.'' \url{https://github.com/Vooban/Smoothly-Blend-Image-Patches},
  Jan. 2022.
\newblock original-date: 2017-08-25T20:26:36Z.

\bibitem{mavrin_focal_2022}
A.~Mavrin, ``Focal {Loss}.'' \url{https://github.com/artemmavrin/focal-loss},
  Jan. 2022.
\newblock original-date: 2019-09-21T05:54:54Z.

\bibitem{taha_metrics_2015}
A.~A. Taha and A.~Hanbury, ``Metrics for evaluating {3D} medical image
  segmentation: analysis, selection, and tool,'' {\em BMC Medical Imaging},
  vol.~15, p.~29, Dec. 2015.

\bibitem{huttenlocher_comparing_1993}
D.~Huttenlocher, G.~Klanderman, and W.~Rucklidge, ``Comparing images using the
  {Hausdorff} distance,'' {\em IEEE Transactions on Pattern Analysis and
  Machine Intelligence}, vol.~15, no.~9, pp.~850--863, 1993.

\bibitem{maiseli_hausdorff_2021}
B.~J. Maiseli, ``Hausdorff {Distance} with {Outliers} and {Noise} {Resilience}
  {Capabilities},'' {\em SN Computer Science}, vol.~2, p.~358, Sept. 2021.

\bibitem{dong-gyu_sim_object_1999}
{Dong-Gyu Sim}, {Oh-Kyu Kwon}, and {Rae-Hong Park}, ``Object matching
  algorithms using robust {Hausdorff} distance measures,'' {\em IEEE
  Transactions on Image Processing}, vol.~8, pp.~425--429, Mar. 1999.

\bibitem{noauthor_surface_2022}
``Surface distance metrics.''
  \url{https://github.com/deepmind/surface-distance}, Jan. 2022.
\newblock original-date: 2018-07-19T13:46:05Z.

\end{thebibliography}

\end{document}